\documentclass[reprint, twocolumn, showkeys,aps,prl,10pt]{revtex4-2}
\usepackage[left=1.25cm,right=1.25cm,top=2.5cm,bottom=2cm]{geometry}
\usepackage[T1]{fontenc}
\usepackage[utf8]{inputenc}
\usepackage[english]{babel}
\usepackage{graphicx}
\usepackage{times}
\usepackage{mathptmx}
\usepackage{eurosym}
\usepackage{amssymb,amsmath,amsfonts}
\usepackage{bm}
\usepackage{latexsym}
\usepackage{balance}
\usepackage{fancyhdr}
\usepackage{MnSymbol}
\usepackage[usenames]{color}
\usepackage{url}
\usepackage{hyperref}
\usepackage{float}
\usepackage{setspace}
\usepackage{balance}

\begin{document}

\title{Condensation energy of superconducting BEC of non-interacting Cooper pairs in multilayers }

\author{I. Ch\'{a}vez}
\email{israelito@fisica.unam.mx}
\author{M.A. Sol\'{i}s}
\email{masolis@fisica.unam.mx}
\address{Instituto de F\'{i}sica, Universidad Nacional Aut\'{o}noma de M\'{e}xico,
Apdo. Postal 20-364, 01000 Mexico City, MEXICO}
\author{P. Salas}
\address{Facultad de Ciencias, Universidad Nacional Aut\'onoma de M\'exico, Apdo. postal 70-542, 04510 Ciudad de México, MEXICO}
\author{J.J. Valencia}
\address{Universidad Autónoma de la Ciudad de México, Plantel San Lorenzo Tezonco,
Apdo. postal 09790, 09780 Ciudad de México, MEXICO}

\begin{abstract}
Boson-Fermion models of superconductivity are getting attention as they are able to explain some of the high temperature superconductor's properties. Here we report on the condensation energy  
of a 3D non-interacting mixture of paired fermions (electrons) as Cooper pairs assumed to be composite bosons, which are responsible for carrying superconductivity, plus unpaired fermions both trapped in a periodic multilayer structure like that of the cuasi-two dimensional High-Temperature superconductor planes, generated by applying an external Dirac’s  comb potential in the direction perpendicular to the planes where superconductivity preferably occurs, while in the other two directions parallel to the planes the mixture moves  freely. For bosons  we give the Bose-Einstein condensation critical temperature, which we assume is equal to the superconducting critical one, while for both bosons and fermions we give the chemical potential, the internal energy and the entropy, all of them as functions of temperature, in order to calculate the Helmholtz free energy, which we use to obtain the condensation energy of a mixture of $N_F$ ideal fermions (electrons), which after turning on the attractive  pair interaction become $N_B = N_F/2$ ideal bosons. 
For several plane impenetrability magnitudes, we calculate the condensation energy as the difference between the free energies of the fermions which are in the normal state minus that of the bosons in the condensed state, where we observe that as the plane impenetrability increases: the condensation energy increases; the critical temperature decreases, as expected for example for cuprate superconductors, and the behavior of the entropy and the internal energy show a dimensional crossover from 3D to 2D.
\end{abstract}

\date{\today}
\keywords{Condensation energy, Bose-Einstein condensation, Boson-Fermion mixtures in multilayers}

\maketitle
%\tableofcontents

\setlength{\parskip}{0pt}
\setcounter{secnumdepth}{1}

\section{Introduction}
\thispagestyle{empty}

Boson-Fermion (BF) models of superconductivity (SC) as a Bose–Einstein condensation (BEC) of Cooper pairs \cite{blatt, shafroth1, shafroth2,shafroth3} date back even before the Bardeen-Cooper-Schrieffer (BCS) theory \cite{BCS} appeared. However, with the advent of high critical temperature superconductors and the inability of BCS theory to explain all their properties, BF theories have gained renewed interest \cite{flee89, flee90, flee91, flee92, casas98, GBEC1, GBEC2}.  Although BCS only addressed the presence of “Cooper correlations” of single-electron states, BF models \cite{raninger85,raninger88,micnas90,micnas2002,micnas2005,GBEC1,GBEC2} postulate the existence of actual bosonic Cooper pairs (CPs). Cooper pairs \cite{Cooper} arise from the interaction of two electrons (or holes) to obtain a bound state of negative (positive) energy with respect to the Fermi surface made up of electrons if their momentum is smaller than the Fermi momentum, $k < k_F$, or holes if $k > k_F$ with an energy exceeding the Fermi energy $E_F$. Here we address those pairs composed by electrons.

Having in mind to arrive to the properties of quasi-two-dimensional high temperature superconductors such as cuprates (YBCO, BiSSCO), in this work we give some thermodynamic properties of the mixture of paired electrons (CPs) plus unpaired electrons within a multilayer structure, where the interactions among them have been turned off once the pairs are formed. The case in which the interactions persist after pairs are formed will be reported elsewhere. The multilayer structure is generated by applying an external Kronig-Penney (KP) potential,  in the limit when barriers become Dirac's deltas, in the direction perpendicular to the planes where superconductivity preferentially occurs. 

The total energy of both bosons or fermions is a sum of the energy in their perpendicular motion ($z$ direction), which satisfies the KP relation \cite{KP}, plus the energy associated with their motion parallel to the planes ($x-y$ directions), which for fermions has a quadratic dispersion relation while Cooper pairs in 2D have a linear-energy-momentum relation as reported in  \cite{adhikari}, which more accurately reproduces the specific heat for cuprates \cite{salas2017}.

In Sec. 2 we introduce the KP relations from which we obtain the energy spectrum for fermions and bosons within the multilayer structure with a separation $a$ between two contiguous planes and for several values of the plane impenetrability. In Sec. 3 we summarize the non-interacting Boson-Fermion model. In Sec. 4, using the grand thermodynamical potential for bosons and fermions in multilayers, given in Appendices B and C, we get their thermodynamic properties, such as the chemical potential, the entropy, the internal energy and the Helmholtz free energy for both gases.
In Sec. 5 we obtain the condensation energy, which is the difference between the Helmholtz free energy of the fermions (electrons) in the normal state and the Helmoltz free energy of the bosons in the condensed state, assuming that all the fermions initially available paired into Cooper pairs at the BEC critical temperature. In Sec. 6 we give our conclusions.
In Appendix A, for completeness, we give the grand thermodynamic potential for the ideal quantum gases.

\section{Multilayer Structure}

We take a BF mixture immersed in a multilayer structure, where the planes are generated  applying a KP potential \cite{KP} in the limit when the barriers become Dirac's deltas, in the perpendicular direction to the planes. The energy for each particle, either bosons or fermions, can be separated into two components $\varepsilon(\kappa) = \varepsilon(\kappa_x,\kappa_y, \kappa_z) = \varepsilon(\kappa_{\rho}) + \varepsilon(\kappa_z)$, where $\varepsilon(\kappa_{\rho})$ corresponds to the  $x-y$ plane and $\varepsilon(\kappa)$ to the $z$-direction. Bosons have a linear-dispersion relation in the $x-y$ plane \cite{adhikari}, while the fermions have a quadratic dispersion relation. In the $z$-direction the energy for each gas satisfies the KP relation
\begin{equation}
\cos(\kappa \cdot a) = \frac{P}{\alpha\cdot a} \sin(\alpha\cdot a) + \cos(\alpha \cdot a)
\label{KPr}
\end{equation}
where $P = m\,a\,V_0/\hbar^2$ is a measure of the plane impenetrability, with $V_0$ the delta strengh of the Dirac Comb potential $v(z) =\sum_{j = -\infty}^{\infty} V_0 \delta(z-j\,a)$, with $(\alpha\,a)^2 = \varepsilon_{\kappa}/(\hbar^2/2m \ a^2)$, $\varepsilon_\kappa$ the energy of each particle. For fermions $\kappa = k_z$ and $m = m_e$ while for bosons $\kappa = K_z$ (the center of mass momentum of the pair in the $z$-direction) and $m = m_b = 2m_e$, with  $m_e$ the electron mass and $a$ the distance between deltas.

\begin{figure}[!htp]
\begin{center}
\includegraphics[width=0.47\textwidth]{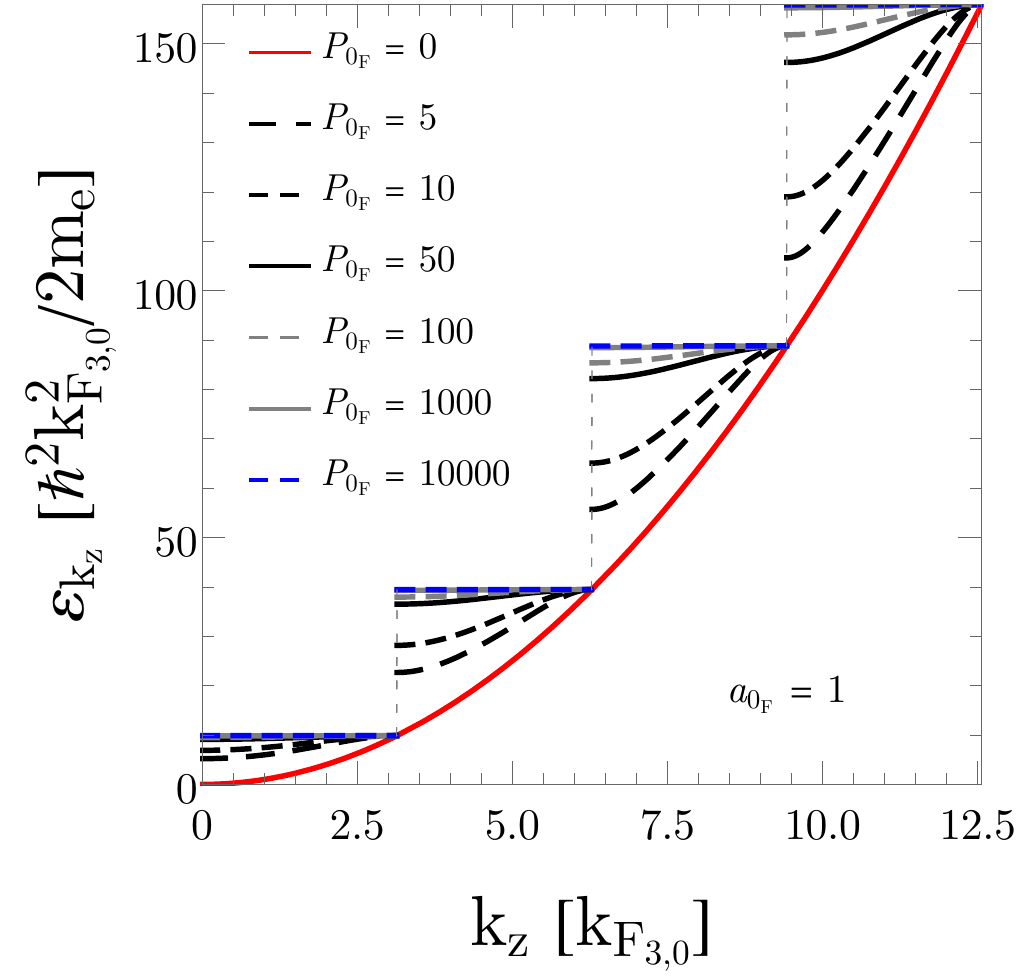}
\end{center}
\caption{Kronig-Penney energy-momentum relation \textit{vs.} $\text{k}_\text{z}$ for a fermion gas using Eq.~\eqref{Eq-KP-adim} changing $P_{0_F}$ and taking $a_{0_F}=1$. The red curve corresponds to the quadratic energy-momentum relation when $P_{0_F}=0$ (ideal case). Note that the energy spectrum tends to a constant as $P_0$ increases for each energy band.}\label{Fig-KP}
\end{figure}

We use the Fermi energy $E_{F_{3,0}}$ and the inverse of the Fermi wave-vector $1/k_{F_{3,0}}$ both of the 3D free IFG as the energy and length units respectively,  where $k_{F_{3,0}} = 3\pi^2 n_{3}^{1/3}$ and $n_3 = N_F/L^3$ the electron number density, with $L^3$ the volume of the system. Using these units, we arrive to the dimensionless KP relation 
\begin{eqnarray}
\cos(\tilde{\kappa}\cdot a_{0_F}) &=& \frac{P_{0_F}}{\tilde{\alpha}} \sin\left(\tilde{\alpha}\cdot a_{0_F} \right)
+ \cos\left( \tilde{\alpha}\cdot a_{0_F} \right) \label{Eq-KP-adim}
\end{eqnarray}
where $a_{0_F} = a~ k_{F_{3,0}}$ and $\tilde{\kappa} = \kappa/k_{F_{3,0}}$ the generalized momentum in units of that of the ideal Fermi gas. Thus for fermions $P_{0_F} = m_e V_0 a_{0_F}/(\hbar^2 k_{F_{3,0}})$ and
$\tilde{\alpha}\cdot a_{0_F} = \tilde{\varepsilon}_{k}^{1/2} a_{0_F}$, with
$\tilde{\alpha} = \alpha/k_{F_0}$,
while for bosons $P_{0_B} = 2P_{0_F}$ and $\tilde{\alpha}_{B} \cdot a_{0_F} = 2^{1/2}\tilde{\varepsilon}_{k}^{1/2} a_{0_F}$, where the subindex $0_F$ refers to the free Fermi energy (not-confined) case. Fig.~\ref{Fig-KP} shows the energy-momentum relation of Eq.~\eqref{Eq-KP-adim} for several values of the strength of the impenetrability $P_{0_F}$ and taking $a_{0_F} =1$. Note that as $P_{0_F}$ increases, the energy relation changes smoothly from a quadratic energy relation to a constant energy relation within each energy band.

The energy spectrum coming from the energy-momentum relation Eq.~\eqref{Eq-KP-adim}, will be inserted in the thermodynamic relations for fermions with a number density $n_F$ and for bosons with a number density $n_B = n_F/2$.

%\balance
\section{Boson-Fermion Model}

We propose a boson-fermion mixture, where  bosons are Cooper-like pairs that appear as resonances of two electrons or two holes as proposed by Friedberg and Lee \cite{FL1, FL2}. The Hamiltonian is
\begin{equation}
H = \sum_{\textbf{k},s} \varepsilon_{\textbf{k}} a_{\textbf{k},s}^{\dagger} a_{\textbf{k},s} + \sum_{\textbf{K}} \varepsilon_{\textbf{K}} b_{\textbf{K}}^{\dagger} b_{\textbf{K}} + H_1
\end{equation}
where $a_{\textbf{k},s}^{\dagger}$ and $b_{\textbf{K}}^{\dagger}$ are the fermion and the composite boson creation operators, respectively, with $s$ the spin and
\begin{equation}
H_1 = \frac{G}{\sqrt{L^3}} \big(a_{\textbf{k}/2 + \textbf{k},s}\, a_{\textbf{k}/2 - \textbf{k},s}\, b_{\textbf{K}}^{\dagger}\; \textsc{v} (\textbf{k}) + h.c. \big)
\end{equation}
the interaction Hamiltonian that creates/destroys composite bosons from/into fermions. Here $\textbf{K} = (K_x, K_y, K_z) \equiv \textbf{k}_1 + \textbf{k}_2$ is the center-of -mass-momentum (CMM) of the pair and $\textbf{k} \equiv (\textbf{k}_1 - \textbf{k}_2)/2$ the relative momentum with $\textbf{k}_1$ and $\textbf{k}_2$ the wave vectors of each electron of the pair. The form factor $\textsc{v}(k)$ is normalized such that $\textsc{v} (0 ) = 1$, which defines the coupling constant $G$. In this version of the model, we assume the zeroth-order approximation \cite{FL1, FL2} so that we keep a mixture of two independent particle systems in a layered structure.

\section{Thermodynamic properties of bosons and fermions}

The thermodynamic properties of the boson and fermion systems immersed in a Dirac's delta comb can be obtained from

\begin{equation}
N \equiv -\left(\frac{\partial \Omega}{\partial \mu} \right)_{T,L^3} \qquad
U \equiv -k_B\,T^2 \frac{\partial}{\partial T} \left(\frac{\Omega}{k_B\,T} \right)_{z, L^3} \label{Rel1}
\end{equation}
\begin{equation}
S \equiv- \left( \frac{\partial \Omega}{\partial T} \right)_{L^{3},\mu}
\qquad
C_V \equiv \left(\frac{\partial U}{\partial T} \right)_{N,L^3} \label{Rel2}
\end{equation}
where the grand thermodynamic potential $\Omega$ for free ideal quantum gases is given in Appendix A, while for bosons and fermions inside multilayers is given in Appendix B and Appendix C, respectively.

The Helmholtz free energy for each system is given by
\begin{equation}
F(T) = U(T) - T\,S(T) \label{Eq-EHelmholtz}
\end{equation}

Thus, the condensation energy is
\begin{equation}
E_c(T) = F_n(T) - F_s(T)
\end{equation}
where $F_n(T) = F_F(T)$ is the Helmholtz free energy of the normal state, i.e., the Helmholtz free energy of the fermion system and $F_s(T) = F_B(T)$ the superconducting Helmholtz free energy, namely, the boson free energy. Taking the number of particles as $N_B = N_F/2$ we solve the system changing $P_{0_F}$ the strength impenetrability and letting $a_{0_F}$ fixed.

\subsection{4.1 \quad Boson Thermodynamic Properties}

In this section we present the main equations of the thermodynamic properties of the boson gas inside the multilayered structure generated by a Dirac's comb potential, using a linear energy-dispersion relation in the $x$-$y$ plane and the KP energy relation in $z$ direction, for several plane impenetrabilities $P_{0_F}$ keeping $a_{0_F}$ fixed. As described in Appendix B, the boson grand thermodynamic potential is
\begin{small}
\begin{align}
\Omega_B &= \Omega_0 -\frac{L^3}{2\,\pi^2} \frac{1}{c^2\,\beta^3}  \sum_{i=1}^{j} \int\limits_{0}^{\pi} dK_{z}\; g_{3}\big[ \exp(-\beta \{e_0 + \alpha_{i}^{2}(K_z) -\mu_B \}) \big] \label{WBose}
\end{align}
\end{small}
\noindent where $e_0 = E_{F_2} + \Delta_0$, $E_{F_2}$ is the Fermi energy in 2D, $\Delta_0$ the superconductor energy gap at zero temperature,  $c=(2/\pi)\hbar v_{F2D}$ is a constant in terms of the Fermi velocity $v_{F2D}$ in 2D  and $\alpha_{i}^{2}(K_z) = \alpha_{i}^{2}(k_z)/2$ comes from the KP relation for bosons.

The critical temperature and the chemical potential as a function of temperature come from solving implicitly the number equation for bosons $N_B$, namely
\begin{align}
N_B &= N_0 + \frac{L^3}{4 \pi^2 c^2} \frac{1}{\beta^2} \int\limits_{-\infty}^{\infty}\; dK_z ~ g_2\left(\exp\{-\beta[e_0 + \varepsilon(K_z)- \mu_B]\}\right), \label{Eq-Bose-N0}
\end{align}
where $N_0$ is the number of bosons in the ground state energy
\begin{equation*}
N_0 =  \frac{\exp \left(-\beta  \left(e_0+\varepsilon _0-\mu_B \right)\right)}{1-\exp \left(-\beta  \left(e_0+\varepsilon _0-\mu_B \right)\right)}.
\end{equation*}
Introducing the KP energy relation 
\begin{equation}
N_B = N_0 + \frac{2\,L^3}{4 \pi^2 c^2} \frac{1}{\beta^2} \sum_{i=1}^{j} \int\limits_{0}^{\pi}\; dK_z ~ g_2\left(z_1 \right) \label{Eq-Bose-N0-KP}
\end{equation}
with $z_1 = \exp\{-\beta[e_0+ \alpha_{i}^{2}(K_z)- \mu_B]$.

\subsubsection{4.1.1 \quad Critical temperature}

\begin{figure}[!ht]
\begin{center}
\includegraphics[width=0.45\textwidth]{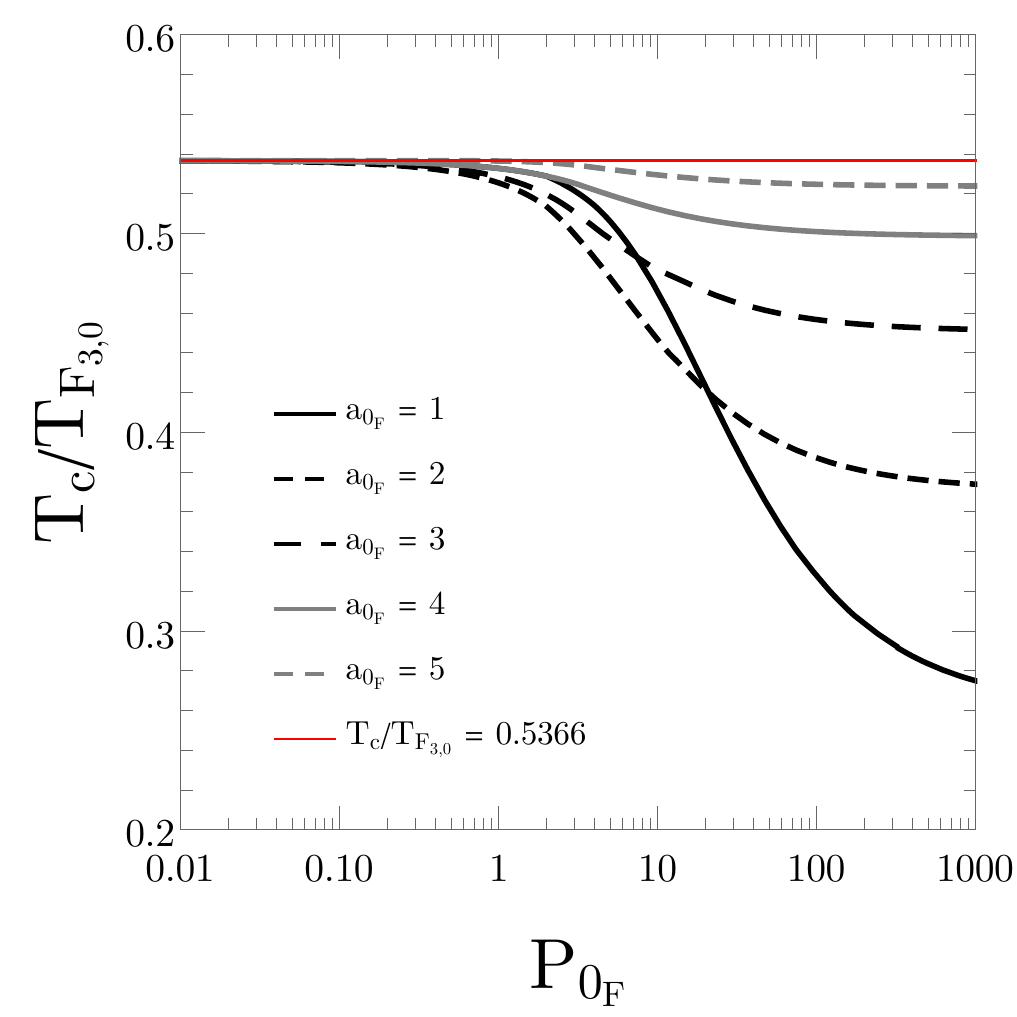}
\end{center}
\caption{Critical temperature of the IBG $T_c/T_{F_{3,0}}$ \textit{vs.} $P_{0_F}$ with $a_{0_F}=1$ (black) curve, $a_{0_F}=2$ (short-dashed-black) curve, $a_{0_F} = 3$ 
 (long-dashed-black) curve, $a_{0_F} = 4$ (gray) curve and $a_{0_F} =5$ (short-dashed-gray) curve. Note that $T_c/T_{F_{3,0}} \to 0.22$, the lowest critical temperature, as $P_{0_F} \to \infty$}. \label{Fig-IBGTcT0}
\end{figure}

In order to get the critical temperature, we take  $T = T_c$, so  $N_0=0$ and the boson chemical potential becomes $\mu_0 = e_0 + \epsilon_0$ with $\varepsilon_0 = \varepsilon[K_z=0,\,P_0,\, a_0]$ the ground state energy, so the critical temperature comes from Eq. \eqref{Eq-Bose-N0-KP} 
\begin{align}
1 &= \frac{L^3}{2\,N_B\,\pi^2 c^2} \frac{1}{\beta_c^2} \sum_{i=1}^{j} \int\limits_{0}^{\pi}\; dK_z ~ g_2\left(\exp\{-\beta_c [\alpha_{i}^{2}(K_z)- \varepsilon_0]\}\right) \label{Eq-BoseTc}
\end{align}
where we have divided by the total number of bosons $N_B$ and $\beta_c = 1/k_B\,T_c$. Fig.\;\ref{Fig-IBGTcT0} shows the critical temperature $T_c/T_{F_{3,0}}$, where $T_{F_{3,0}}$ is the Fermi temperature in 3D for a gas of \textit{free} fermions, as a function of the impenetrability  $P_{0_F}$ for different values of the separation between planes $a_{0_F}=1, 2, 3, 4, 5$. Taking $a_{0_F}=1$, when $P_{0_F} \to \infty$, i.e., the 3D system tends to a cuasi-2D system, the BEC critical temperature value $T_c/T_{F_{3,0}} \to 0.22$, which is $a_{0_F}$ value dependent.
Also, the critical temperature decreases as the strength impenetrability increases with the distance between planes fixed.

\subsubsection{4.1.2 \quad Chemical potential}

\begin{figure}[!h]
\begin{center}
\includegraphics[width=0.45\textwidth]{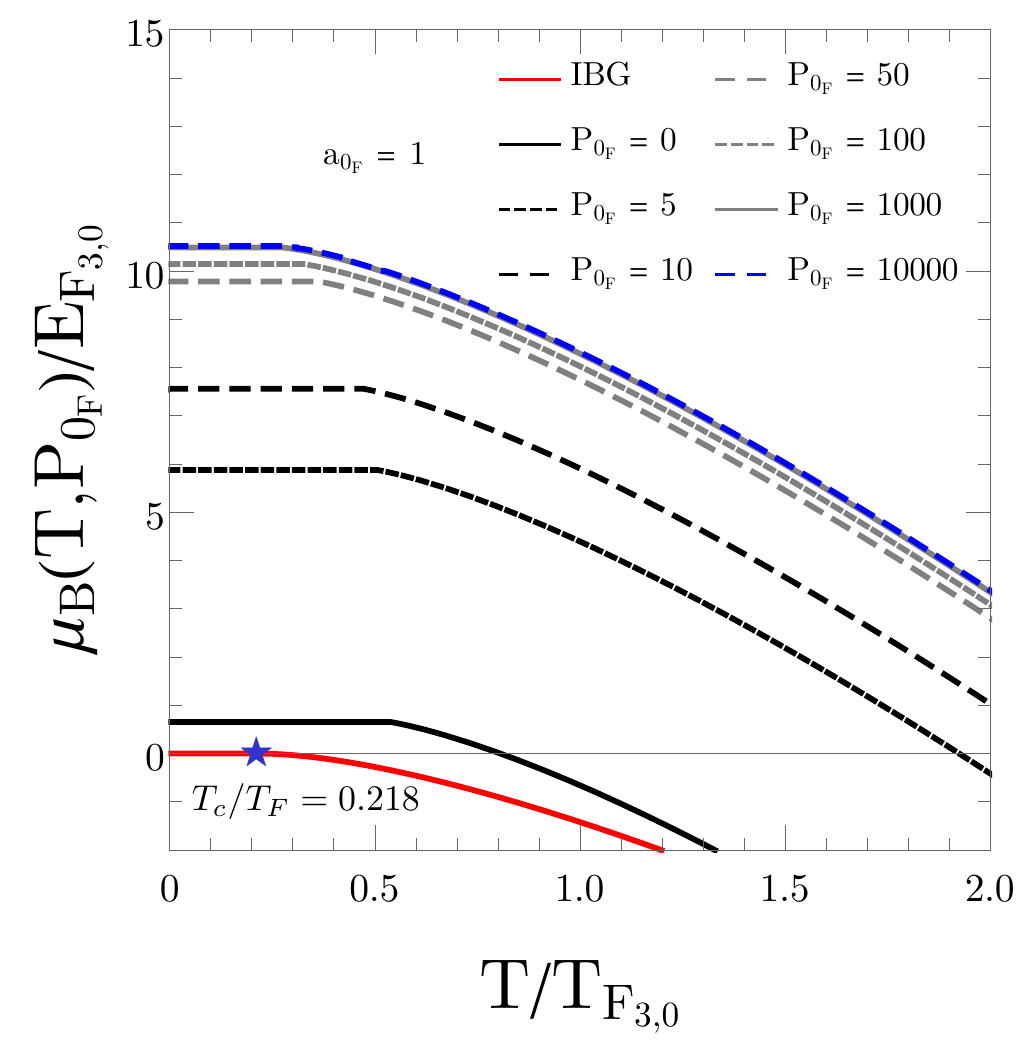}
\caption{Chemical potential of the boson gas $\mu_B(T,P_{0_F})/E_{F_{3,0}}$ \textit{vs.} $T/T_{F_{3,0}}$ for different values of $P_{0_F}$ and with $a_{0_F}=1$. For comparison purposes we show the chemical potential of the free IBG (red) curve, with $T_c/T_{F_{3,0}} = 0.218$ the critical temperature of the IBG (blue star).}\label{Fig-IBGmu}
\end{center}
\end{figure}

We find the chemical potential for this boson system when $T> T_c$ from Eq. \eqref{Eq-Bose-N0}, so
\begin{flalign*}
%\begin{eqnarray}
1 &= \frac{1}{4\,n_B\,\pi^2\,(\beta c)^2} \int_{-\infty}^{\infty} dK_z~ g_{2}(\exp[-\beta \{e_0 + \varepsilon(K_z) - \mu_B \}]), %\label{Eq-IBGmu}
%\end{eqnarray}
\end{flalign*}
and using the KP relation
\begin{flalign}
%\begin{eqnarray}
1 &= \frac{1}{2\,n_B\,\pi^2\,(\beta c)^2} \sum_{i=1}^{j} \int\limits_{0}^{\pi} dK_z~ g_{2}(\exp[-\beta \{e_0 + \alpha_{i}^{2}(K_z)/2 - \mu_B \}]) \label{Eq-IBGmu-KP}
%\end{eqnarray}
\end{flalign}
Fig.\,\ref{Fig-IBGmu} shows the chemical potential of this boson system for $T \geq 0$. As shown before, the critical temperature decreases as the strength impenetrability $P_{0_F}$ increases. Also note that the ground state energy, i.e., the boson chemical potential $\mu_{B}(T=0)$ increases continuously but with a decreasing magnitude as $P_{0_F}$ increases, where the zero value corresponds to the free case. This behavior will be reflected in the boson Helmholtz free energy.

\subsubsection{4.1.3 \quad Entropy}

The entropy of the boson system in multilayers is given by
%\begin{eqnarray*}
\begin{align*}
S_B 
%&\equiv \left( \frac{\partial \Omega_B}
%%{\partial T} \right)_{L^{3},\mu_B}
%\notag \\
&= \frac{L^3}{4 \pi ^2 c^2}\frac{k_B}{\beta} \int\limits_{-\infty }^{\infty } dK_z 
\left( e_0 + \varepsilon(K_z)-\mu_B \right)~
\notag \\
&\times g_2\left(\exp\{-\beta [\varepsilon(K_z)+ e_0 -\mu_B] \} \right)
\notag \\
&+ \frac{3 L^3}{4 \pi ^2 c^2} \frac{k_B}{\beta^2} \int\limits_{-\infty }^{\infty }dK_z~ g_3\left(\exp\{-\beta [\varepsilon(K_z)+ e_0 -\mu_B] \} \right)
%\end{eqnarray*}
\end{align*}
which after dividing by $N_Bk_B$ and introducing the KP energy relation becomes
\begin{align}
\frac{S_B}{N_B\,k_B} &= \frac{L^3}{2\,N_B \pi ^2 c^2}\frac{1}{\beta} \sum_{i=1}^{j} \int\limits_{0}^{\pi} dK_z 
\left( e_0 + \alpha_{i}^{2}(K_z)-\mu_B \right)
\notag \\
&\times g_2\left(\exp\{-\beta [\alpha_{i}^{2}(K_z)+ e_0 -\mu_B] \} \right)
\notag \\
&+ \frac{3\, L^3}{2\,N_B \pi ^2 c^2} \frac{1}{\beta^2} \sum_{i=1}^{j}\int\limits_{0}^{\pi}dK_z~ g_3\left(\exp\{-\beta [\alpha_{i}^{2}(K_z)+ e_0 -\mu_B] \} \right), \label{Eq-SB-KP}
\end{align}
where $k_B$ is the Boltzmann constant.

\begin{figure}[!htbp]
\begin{center}
\includegraphics[width=0.44\textwidth]{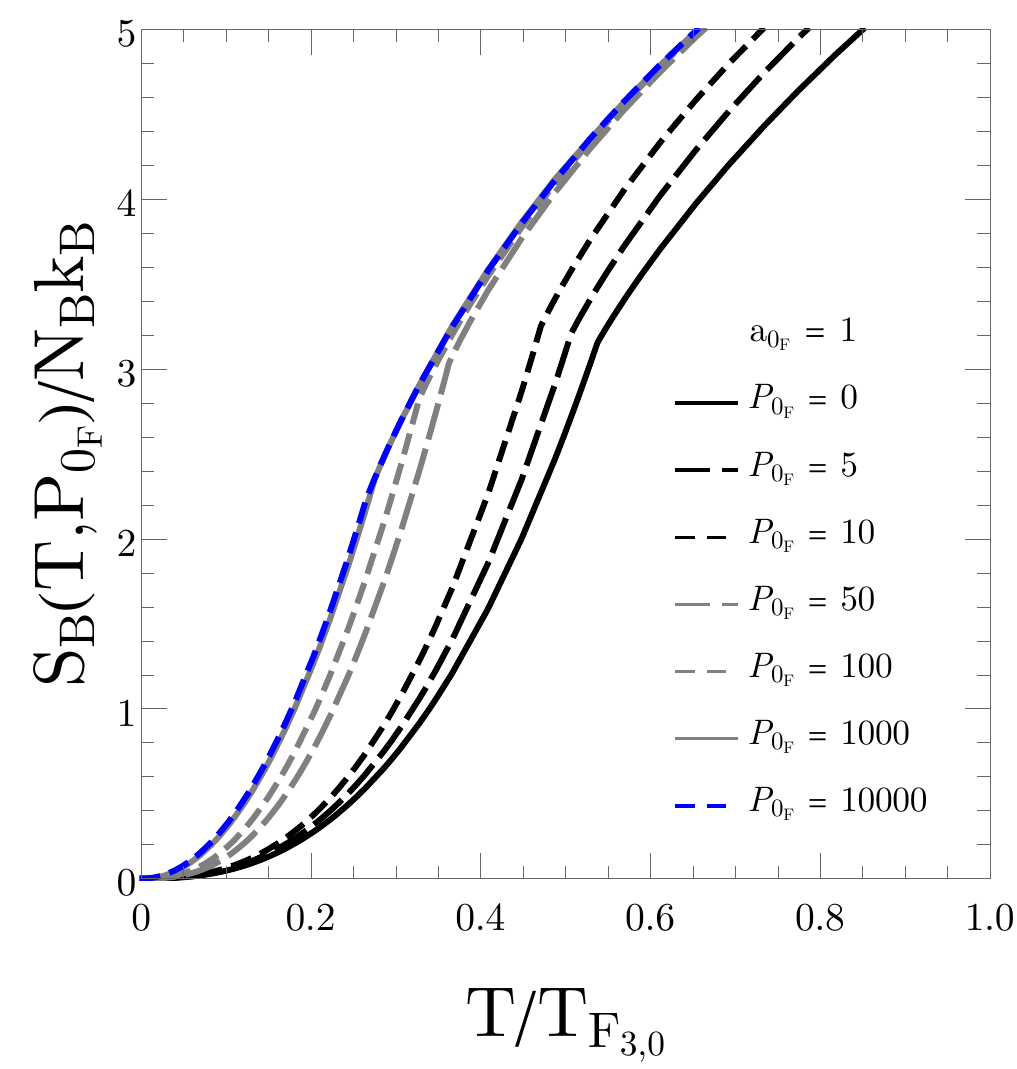}
\end{center}
\caption{Entropy  $S_B(T,P_{0_F})/N_B\,k_B$ \textit{vs.} $T/T_{F_{3,0}}$ of a boson gas inside multilayers for several values of $P_{0_F}$ and $a_{0_F}=1$.} \label{Fig-SB-KP}
\end{figure}

In Fig.~\ref{Fig-SB-KP} we plot Eq. \eqref{Eq-SB-KP}  for several values of $P_{0_F}$ and $a_{0_F} = 1$. Note that for each curve there is a sudden change in the curvature at the corresponding critical temperature, indicating a phase transition.
%\vspace*{1cm}

%\vspace*{-0.95cm}
\subsubsection{4.1.4 \quad Internal energy}

\begin{figure}[!ht]
\begin{center}
\includegraphics[width=0.45\textwidth]{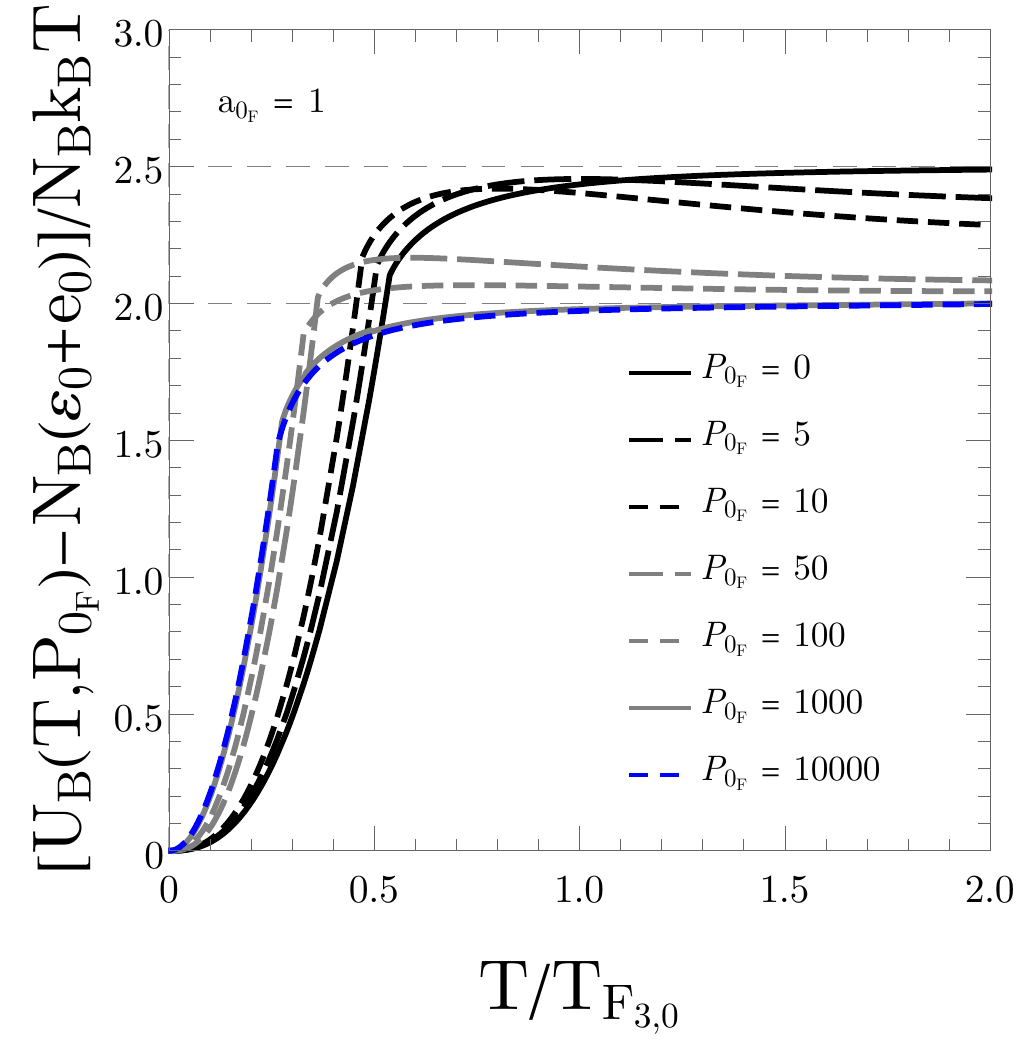}
\end{center}
\caption{Internal energy $[U_B(T,P_{0_F})-N_{B}\,(e_{0} + \varepsilon_{0})]/N_{B}\,k_{B}T$ \textit{vs.} $T/T_{F_{3,0}}$ changing $P_{0_F}$ with $a_{0_F} =1$. Thin-dashed horizontal lines mark the limits of the internal energy as $T/T_{F_{3,0}} \to \infty$.}
\label{Fig-Uint}
\end{figure}

The internal energy of this boson system comes from Eq. \eqref{Rel1} as
\begin{flalign*}
U &= U_0 + \frac{L^3}{4 \pi ^2 c^2} \frac{1}{\beta^2} \int\limits_{-\infty }^{\infty } dK_z \left[ e_0+\varepsilon(K_z) \right]
\notag \\
&\times g_2\left(\exp\{-\beta [\varepsilon(K_z)+e_0-\mu_B] \}\right)
\notag \\
&+ \frac{L^3}{2 \pi ^2 c^2} \frac{1}{\beta ^3} \int\limits_{-\infty }^{\infty } dK_z ~ g_3\left(\exp \{-\beta [\varepsilon(K_z) + e_0 - \mu_B] \} \right) 
\end{flalign*}
where
\begin{equation}
U_0 = \frac{\left(e_0+\varepsilon _0\right) \exp \left(-\beta  \left(e_0+\varepsilon _0-\mu_B \right)\right)}{1-\exp \left(-\beta  \left(e_0+\varepsilon _0-\mu_B \right)\right)} = \left(e_0+\varepsilon _0\right)N_0 \notag
\end{equation}
is the ground state energy. After adding and subtracting $\varepsilon_0\,N_e$ on the right side of the previous equation, and using $(e_0 + \varepsilon_0)N_B = (e_0 + \varepsilon_0)\,N_0  + (\varepsilon_0 + e_0)\,N_e$, we have
\begin{small}
\begin{flalign}
U - (e_0 + \varepsilon_0)N_B &= \frac{L^3}{4 \pi ^2 c^2} \frac{1}{\beta^2} \int\limits_{-\infty }^{\infty } dK_z [\varepsilon(K_z)-\varepsilon_0]
\notag \\
&\times  g_2\left(\exp\{-\beta [e_0 + \varepsilon(K_z)- \mu_B] \}\right)
\notag \\
&+ \frac{L^3}{2 \pi ^2 c^2} \frac{1}{\beta ^3} \int\limits_{-\infty }^{\infty } dK_z ~ g_3\left(\exp \{-\beta [e_0 + \varepsilon(K_z) - \mu_B] \}\right) \label{Eq-Cv01}
\end{flalign}
\end{small}

which dividing by $N_B\,k_B\,T$ and introducing the KP relations gives
\begin{align}
%\begin{eqnarray}
\frac{U - (e_0 + \varepsilon_0)N_B}{N_B\,k_B\,T} &= \frac{L^3}{2\,N_B\, \pi ^2 c^2} \frac{1}{\beta} \sum_{i=1}^{j} \int\limits_{0}^{\pi}~ dK_z~ [\alpha_{i}^{2}(K_z)-\varepsilon_0]
\notag \\
&\times
g_2\left(\exp\{-\beta [e_0 + \alpha_{i}^{2}(K_z) - \mu_B] \}\right)
\notag \\
&+ \frac{L^3}{N_B\, \pi ^2 c^2} \frac{1}{\beta^2} \sum_{i=1}^{j} \int\limits_{0}^{\pi} dK_z
\notag \\
&\times g_3\left(\exp \{-\beta [e_0 + \alpha_{i}^{2}(K_z) - \mu_B] \}\right) \label{Eq-UIB-KP}
%\end{eqnarray}
\end{align}
In Fig.~\ref{Fig-Uint} we show the internal energy over the temperature $T$ for a Bose gas inside Dirac's deltas using Eq. \eqref{Eq-UIB-KP}, for several plane impenetrabilities $P_{0_F}$ and for an $a_{0_F}$ fixed. When $P_{0_F}=0$ the internal energy over the temperature tends to $2.5$, the 3D classic limit, as $T/T_{F_{3,0}}\to \infty$, while when $P_{0_F}$ increases the limit tends to $2$, the corresponding 2D value.

\subsubsection{4.1.5 \quad Helmholtz free energy}

\begin{figure}[!htbp]
\begin{center}
\includegraphics[width=0.47\textwidth]{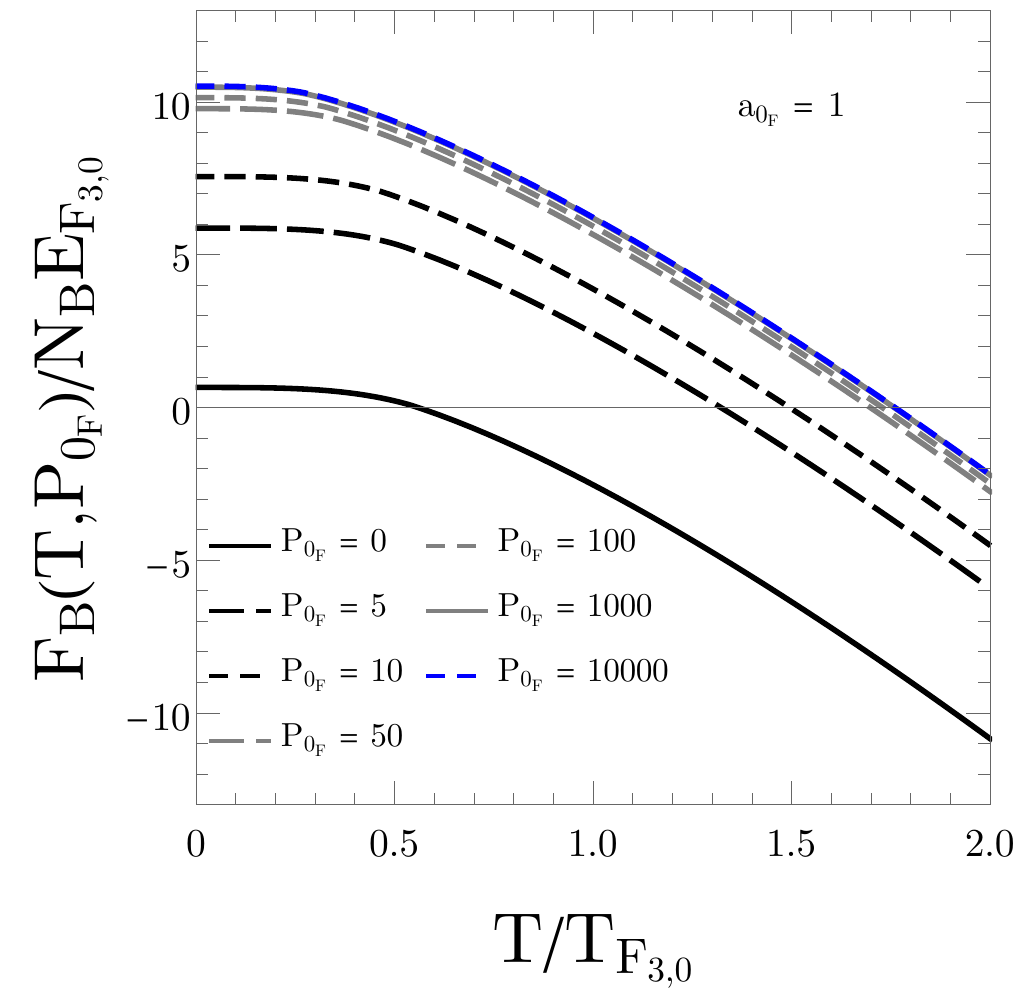}
\end{center}
\caption{Boson Helmholtz free energy inside multilayers  $F_B(T,P_{0_F})/N_B\,E_{F_{3,0}}$ \textit{vs.} $T/T_{F_{3,0}}$ for several values of $P_{0_F}$ with $a_{0_F}=1$. Note that $F_{B}(T=0)$ increases as $P_{0_F}$ increases.}\label{Fig-FB-KP}
\end{figure}

Now we are able to calculate the Helmholtz free energy for the boson system inside a multilayered structure, namely
\begin{equation}
F_B(T) = U_B(T) - T\,S_B(T)
\end{equation}
%\vspace*{-0.5cm}
thus
\begin{flalign*}
F_B &=  (e_0 + \varepsilon_0)N_B + \frac{L^3}{4\, \pi^2 c^2} \frac{1}{\beta^2} \int\limits_{-\infty }^{\infty } dK_z [\varepsilon(K_z)-\varepsilon_0]
\notag \\
&\times g_2\left(\exp\{-\beta [e_0 +\varepsilon(K_z) -\mu_B] \}\right)
\notag \\
&+ \frac{L^3}{2\, \pi ^2 c^2} \frac{1}{\beta^3} \int\limits_{-\infty }^{\infty } dK_z ~ g_3\left(\exp \{-\beta [e_0 +\varepsilon(K_z) - \mu_B] \}\right)
\notag \\
&-  \frac{L^3}{4\, \pi ^2 c^2}\frac{1}{\beta^2} \int\limits_{-\infty }^{\infty } dK_z \left( e_0 + \varepsilon(K_z)-\mu_B \right)~
\notag \\
&\times g_2\left(\exp\{-\beta [e_0 +\varepsilon(K_z) -\mu_B] \} \right)
\notag \\
&- \frac{3 L^3}{4\, \pi ^2 c^2} \frac{k_B}{\beta^3} \int\limits_{-\infty }^{\infty }dK_z~ g_3\left(\exp\{-\beta [e_0 +\varepsilon(K_z) -\mu_B] \} \right)
\end{flalign*}
which after some algebra and introducing the KP relation gives
%\vspace*{-1cm}
\begin{small}
\begin{flalign}
\frac{F_B}{N_B\,k_B\,T} &= \beta (e_0+\varepsilon_0) + \frac{1}{2\,n_B \pi ^2 c^2} \frac{1}{\beta} \sum_{i=1}^{j} \int\limits_{0}^{\pi}\; dK_z ~ \left(e_0 -\alpha_{i}^{2}(K_z)-\mu_B \right)
\notag \\
&\times g_2\left(\exp\{-\beta [\alpha_{i}^{2}(K_z)+e_0-\mu_B] \}\right)
\notag \\
&- \frac{1}{2\,n_B\, \pi ^2 c^2} \frac{1}{\beta^2} \sum_{i=1}^{j} \int\limits_{0}^{\pi}\; dK_z ~ g_3\left(\exp\{-\beta [e_0 + \alpha_{i}^{2}(K_z) -\mu_B] \} \right)
. \label{Eq-FB-KP}
\end{flalign}
\end{small}
In terms of the Fermi energy the last equation becomes
\begin{small}
\begin{flalign}
\frac{F_B}{N_B\,k_B\,T_F} &= \frac{e_0+\alpha_{i}^{2}(K_z=0)}{k_B\,T_F}
\notag \\
&+ \frac{1}{2\,n_B \pi ^2 c^2}\frac{1}{\beta} \sum_{i}^{j} \int\limits_{0}^{\pi}\; dK_z ~ \left(\frac{e_0-\alpha_{i}^{2}(K_z=0) -\mu_B}{k_B\,T_F} \right)
\notag \\
&\times g_2\left(\exp\{-\beta [\alpha_{i}^{2}(K_z)+e_0-\mu_B] \}\right)
\notag \\
&- \frac{1}{4\,n_B\,k_B\,T_F\, \pi ^2 c^2} \frac{1}{\beta ^2} \int\limits_{-\infty }^{\infty }\; dK_z\, g_3\left(\exp\{-\beta [e_0 +\alpha_{i}^{2}(K_z) -\mu_B] \} \right) \label{Eq-FB-EF-KP}
\end{flalign}
\end{small}
Fig.~\ref{Fig-FB-KP} shows the Helmholtz free energy of this boson system inside multilayers, namely $F_B(T,P_{0_F})/N_B\,E_{F_{3,0}}$, with a linear dispersion relation in the $x$-$y$ plane and a KP relation in the perpendicular direction to that plane, taking different values for the strength of the deltas $P_{0_F}$ and $a_{0_F} = 1$. 
As we mentioned before, the ground state Helmholtz energy increases continuously but with a decreasing magnitude  as the  impenetrability strength increases.
%
%\vspace*{-0.5cm}

\subsection{4.2 \quad Fermion Thermodynamic Properties}

For fermions, the energy-momentum relation is quadratic in the direction of the planes and with a KP potential in the perpendicular direction, as described in Appendix C, thus the thermodynamic potential gives
\begin{align}
\Omega_F &= -\frac{L^3}{\pi^2} \left(\frac{m_e}{\beta^2\,\hbar^{2}}\right) \sum_{i=1}^{j}\int_{0}^{\pi} dk_z\; f_{2}\big[\exp[-\beta\,(\alpha_{i}^{2}(k_z)-\mu_F)]\big]. \label{WFermi}
\end{align}

In this section we present the main equations of the thermodynamic properties for the fermion gas, where the particles are free to move with a quadratic energy dispersion relation in the $x$-$y$ direction and in the $z$-direction they are inside the same Dirac's comb potential as before, changing the strength of delta potentials $P_{0_F}$ and  $a_{0_F}$ fixed. As we mentioned earlier, by increasing $P_{0_F}$ we force the system to go from a 3D  to a quasi-2D system.

%\vspace*{-1cm}
\subsubsection{4.2.1 \quad Number of particles}

\begin{figure}[!htp]
\begin{center}
\includegraphics[width=0.45\textwidth]{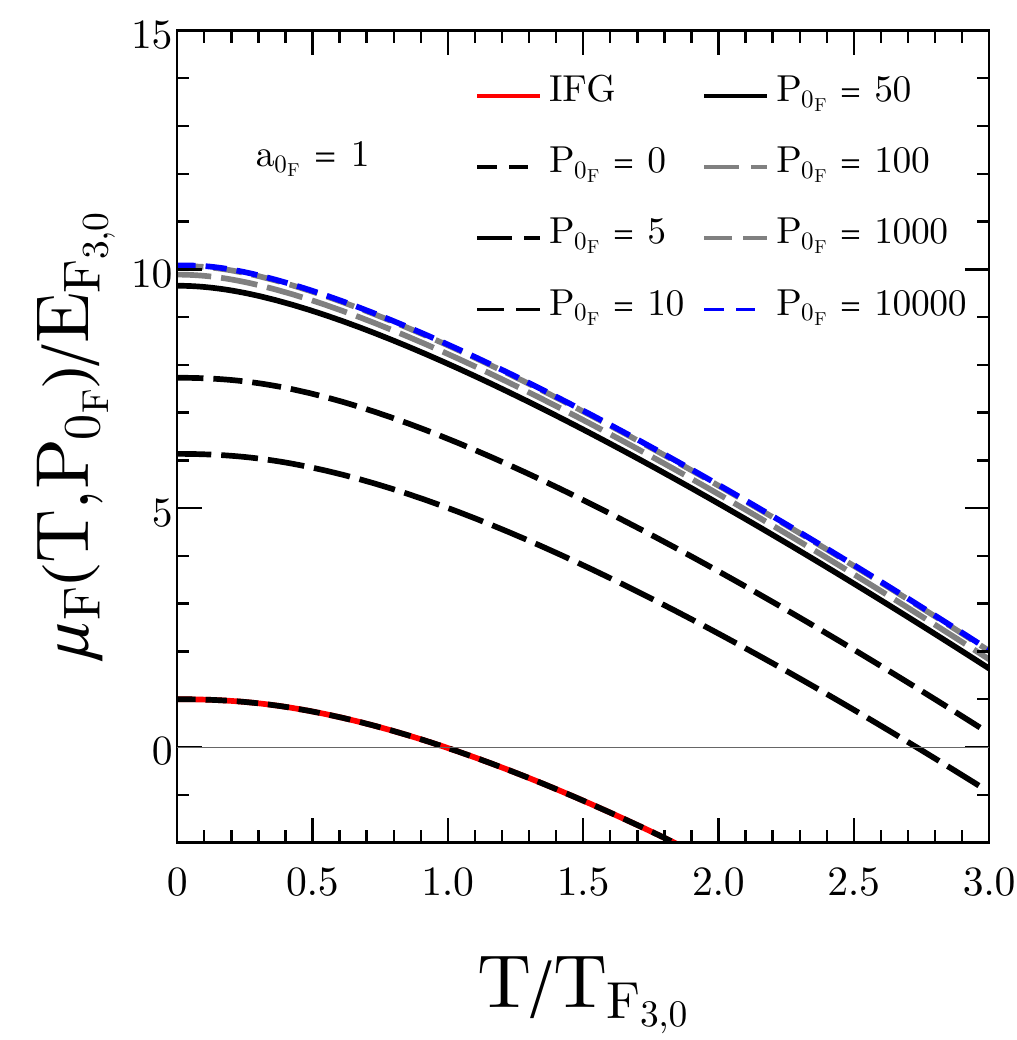}
\end{center}
\caption{Fermion chemical potential $\mu_F(T,P_{0_F})/E_{F_{3,0}}$ \textit{vs.} $T/T_{F_{3,0}}$ as function of the strength of the deltas $P_{0_F}$ and the separation $a_{0_F}=1$. Here $k_B\,T_{F_{3,0}} = E_{F_{3,0}}$ is the energy of IFG system in 3D. Note that the $P_{0_F}=0$ (black-dashed) curve coincides with the IFG (red) curve.}\label{Fig-FChempot}
\end{figure}

From the first relation in \eqref{Rel1}, and  using Eq. \eqref{WFermi} one can find the number of particles of the fermion system as
\begin{eqnarray*}
N_F &=& \left(\frac{L^3}{2\pi^2} \right) \left(\frac{m_e}{\hbar^2 \,\beta} \right) \int_{-\infty}^{\infty} \,dk_z \ln \big[1+\exp\{-\beta (\varepsilon(k_z) - \mu_F) \} \big] \label{Eq-NF}
\end{eqnarray*} 
and using the KP energy relation, gives
%\begin{eqnarray}
\begin{align}
N_F &= \left(\frac{L^3}{\pi^2} \right) \left(\frac{m_e}{\hbar^2 \,\beta} \right) \sum_{i=1}^{j} \int_{0}^{\pi} \,dk_z \ln \big[1+\exp\{-\beta (\alpha_{i}^{2}(k_z) - \mu_F) \} \big]. \label{Eq-NF-KP}
\end{align}
%\end{eqnarray} 
The chemical potential $\mu_F(T,P_{0_F})/E_{F_{3,0}}$ for fermions with a quadratic energy-momentum relation can be obtained from the last equation as a function of the absolute temperature and the strength of the delta potentials, which is shown in Fig.\ref{Fig-FChempot}. Note that the chemical potential with $P_{0_F}=0$ (black-dashed) curve coincides with that of the IFG (red) curve, as expected. As $P_{0_F}$ increases the ground state energy of the system also increases.

\subsubsection{4.2.2 \quad Entropy}

The entropy for the fermion system inside multilayers is given by
\begin{align}
S_F 
%&\equiv - \left( \frac{\partial \Omega_F}
%{\partial T} \right)_{L^{3},\mu_F}
%\notag \\
&= \left(\frac{L^{3}k_B}{2\,\pi^{2}} \right) \left(\frac{m_e}{\hbar^2} \right) \int\limits_{-\infty}^{\infty} dk_z ~ (\varepsilon(k_z) - \mu_F)
\notag \\
&\times \ln \big[1+\exp \{\beta (\varepsilon(k_z) - \mu_F) \} \big]
\notag \\
&+ \left(\frac{L^{3} k_B}{2\,\pi^{2}} \right) \left(\frac{m_e}{\hbar^2\,\beta} \right) \int\limits_{-\infty}^{\infty} dk_z ~ f_2\big[-\exp\{-\beta (\varepsilon(k_z) - \mu_F) \} \big], \label{Eq-SF}
\end{align}

which dividing both sides by $N_F\,k_B$ and using the KP energy relations gives

%\begin{eqnarray}
\begin{align}
\frac{S_F(T)}{N_F\,k_B}&= \left(\frac{L^{3}}{N_F\,\pi^{2}} \right) \left(\frac{m_e}{\hbar^2} \right) \sum_{i=1}^{j} \int\limits_{0}^{\pi} dk_z (\alpha_{i}^{2}(k_z) - \mu_F)
\notag \\
&\times  \ln \big[1+\exp \{\beta (\alpha_{i}^{2}(k_z) - \mu_F) \} \big]
\notag \\
&+ \left(\frac{L^{3}}{N_F\,\pi^{2}} \right) \left(\frac{m_e}{\hbar^2\,\beta} \right) \sum_{i=1}^{j} \int\limits_{0}^{\pi} dk_z~ f_2\big[-\exp\{-\beta (\alpha_{i}^{2}(k_z) - \mu_F) \} \big]. \label{Eq-SF-KP}
\end{align}
%\end{eqnarray}

\begin{figure}[!htp]
\begin{center}
\includegraphics[width=0.44\textwidth]{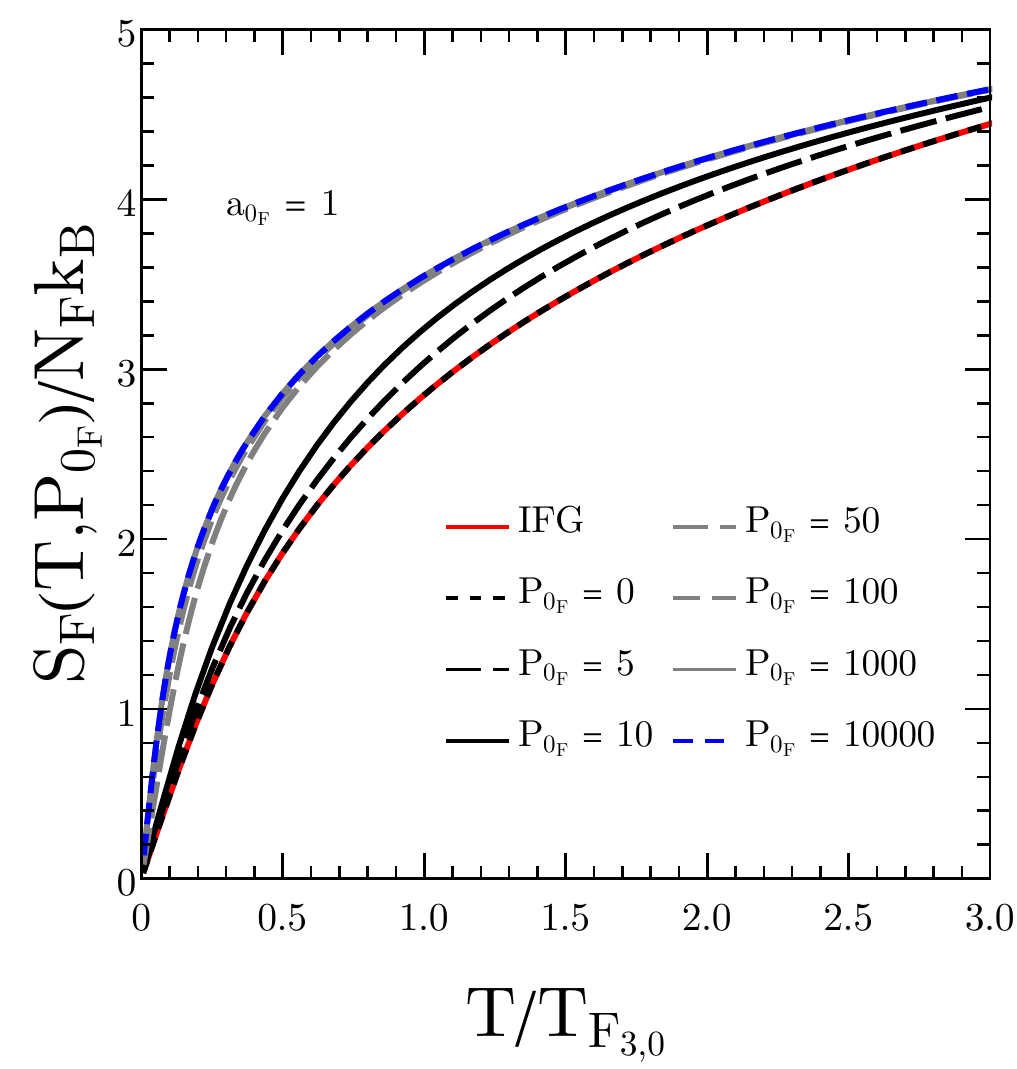}
\end{center}
\caption{Fermion entropy $S_F(T,P_{0_F})/N_F\,k_B$ \textit{vs.} $T/T_{F_{3,0}}$ for different values of $P_{0_F}$ with $a_{0_F}=1$. For comparison, the IFG (red) curve  is plotted, which coincides with $P_{0_F} =0$ (black-dashed) curve.}\label{Fig-S-KP}
\end{figure}

The last equation is plotted in Fig.\ref{Fig-S-KP} for different values of $P_{0_F}$ with $a_{0_F}=1$. Note that as the strength of the potentials increases the slope of the curves at low temperatures also increases.

\subsubsection{4.2.3 \quad Internal energy}

The internal energy for the fermion gas inside multilayers can be calculated from
\begin{eqnarray*}
U_F &=& \left(\frac{L^3}{2\pi^{2}} \right) \frac{m_e}{\hbar^2\,\beta} \int\limits_{-\infty}^{\infty} dk_z \varepsilon(k_z) \ln \big[1+\exp\{-\beta(\varepsilon(k_z) - \mu_F) \} \big]
\notag \\
&+& \left(\frac{L^3}{2\pi^{2}} \right) \frac{m_e}{\hbar^2\,\beta^{2}} \int\limits_{-\infty}^{\infty} dk_z~ f_{2} \big[-\exp\{-\beta(\varepsilon(k_z)-\mu_F) \} \big].
\end{eqnarray*}
\break
dividing both sides by $N_F\,k_B\,T_F$ and once again using the KP energy relations, gives

%\begin{small}
\begin{flalign}
\frac{U_F}{N_F\,k_B\,T} &= \left(\frac{L^3\,m_e}{N_F\,\pi^{2}\,\hbar^2} \right) \frac{1}{\beta} \sum_{i=1}^{j} \int\limits_{0}^{\pi} dk_z ~\alpha_{i}^{2}(k_z)
\notag \\
&\times \ln \big[1+\exp\{-\beta(\alpha_{i}^{2}(k_z) - \mu_F) \} \big]
\notag \\
&+ \left(\frac{L^3\,m_e}{N_F\,\pi^{2}\,\hbar^2} \right) \frac{1}{\beta^2} \sum_{i=1}^{j} \int\limits_{0}^{\pi} dk_z~ f_{2} \big[-\exp\{-\beta(\alpha_{i}^{2}(k_z)-\mu_F)\} \big]. \label{Eq-UF-KP}
\end{flalign}
%\end{small} 

\begin{figure}[!htp]
\begin{center}
\includegraphics[width=0.46\textwidth]{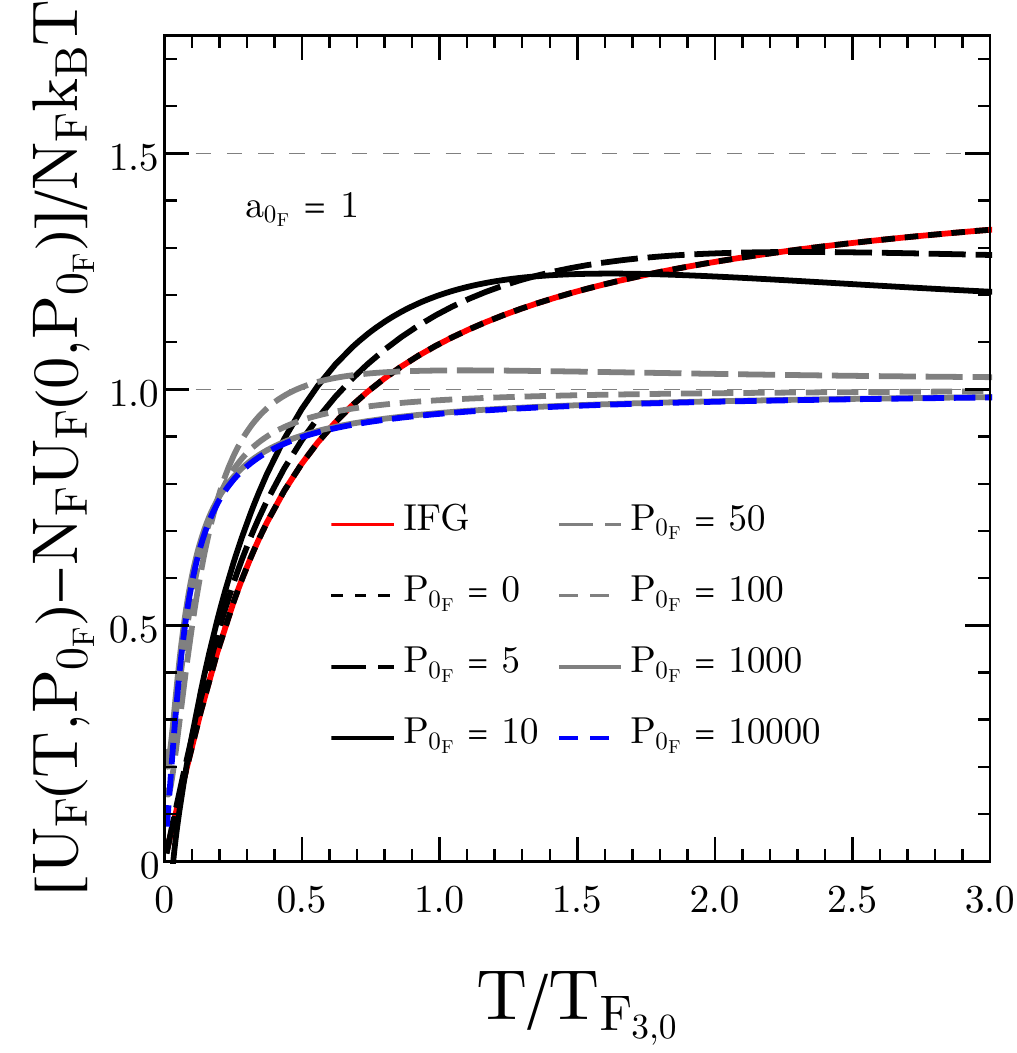}
\end{center}
\caption{Fermion internal energy $[U_F(T,P_{0_F})-N_F\,U_F(0,P_{0_F})]/N_F\,k_B\,T$ \textit{vs.} $T/T_{F_{30}}$ of this fermion system into multilayers, changing the strength of the deltas $P_{0_F}$ with $a_{0_F}=1$. Once again, the internal energy of IFG (red) curve, coincides  with that of $P_{0_F}=0$ (black-dashed) curve.}\label{Fig-UF-KP}
\end{figure}

In Fig.~\ref{Fig-UF-KP} we plot Eq.~\eqref{Eq-UF-KP} subtracting the quantity $U_0/k_B\,T \equiv U_F(T=0,P_{0_F})/k_B\,T$ on both sides of the equation. For comparison purposes the internal energy of the IFG is also plotted. The quantity $U_F(T=0,P_{0_F})/k_B\,T_F$ increases as $P_{0_F}$ increases, although this behavior it is not shown in that figure.

%\vspace*{-1cm}

%\vspace*{-2.5cm}
\newpage
\subsubsection{4.2.4 \quad Helmholtz Free Energy}

\begin{figure}[!ht]
\vspace{0.5cm}
\begin{center}
\includegraphics[width=0.47\textwidth]{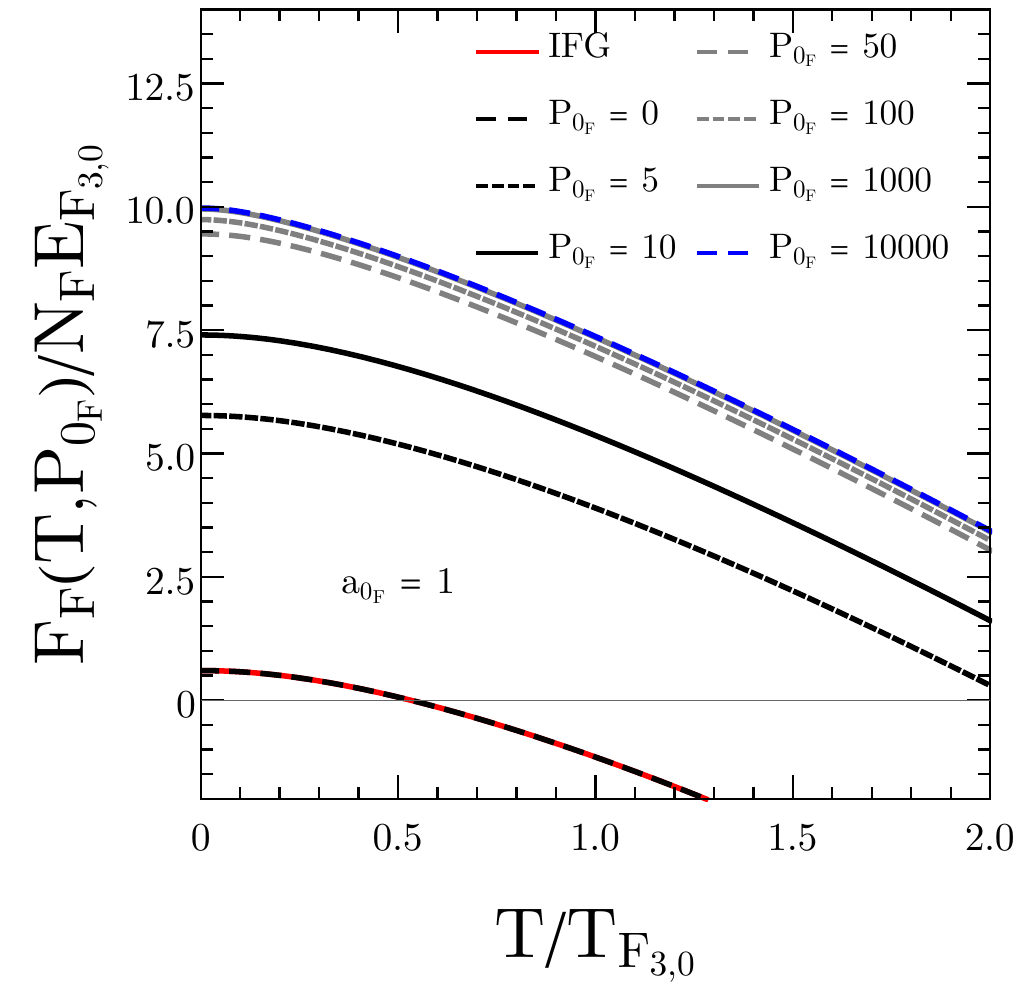}
\end{center}
\caption{Helmholtz free energy \textit{vs.} $T/T_{F_{3,0}}$ for the fermion system changing $P_{0_F}$ with $a_0=1$. Note once again that the ground state energy increases as $P_{0_F}$ increases.}\label{Fig-HF-KP}
\end{figure}

The Helmholtz free energy is $F = U - T S$, so for this fermionic system inside a multilayered structure, and using the previously obtained properties
%\begin{small}
\begin{eqnarray*}
F_F(T) &\equiv & U_F(T) - T S_F(T)
\notag \\
&=& \left(\frac{L^3}{2\pi^{2}} \right) \frac{m_e}{\hbar^2\,\beta} \int\limits_{-\infty}^{\infty} dk_z \varepsilon(k_z) \ln \big[1+\exp\{-\beta(\varepsilon(k_z) - \mu_F) \} \big]
\notag \\
&-& \left(\frac{L^3}{2\pi^{2}} \right) \frac{m_e}{\hbar^2\,\beta^{2}} \int\limits_{-\infty}^{\infty} dk_z f_{2} \big[-\exp\{-\beta(\varepsilon(k_z)-\mu_F)\} \big]
\notag \\
&-&\left(\frac{L^{3}\,T\,k_B}{2\,\pi^{2}} \right) \left(\frac{m_e}{\hbar^2} \right) \int\limits_{-\infty}^{\infty} dk_z ~ (\varepsilon(k_z) - \mu_F)
\notag \\
&\times & \ln \big[1+\exp \{\beta (\varepsilon(k_z) - \mu_F) \} \big]
\notag \\
&-& \left(\frac{L^{3}\,T\, k_B}{2\,\pi^{2}} \right) \left(\frac{m_e}{\hbar^2\,\beta} \right) \int\limits_{-\infty}^{\infty} dk_z ~ f_2\big[-\exp\{-\beta (\varepsilon(k_z) - \mu_F) \} \big]. \qquad
\end{eqnarray*}
%\end{small}
Dividing both sides by $N_F\,k_B\,T_F$, reducing terms and introducing the KP energy relations
%\vspace{1cm}
%\begin{small}
\begin{align}
\frac{F}{N_F\,k_B\,T_F} &= \frac{L^3}{N_F \pi ^2} \frac{m_e}{\hbar ^2} \frac{1}{\beta} \sum_{i=1}^{j} \int\limits_{-0}^{\pi} \;dk_z~ \mu_F 
\notag \\
&\times \ln \left(1+ \exp \left(-\beta  \left(\alpha_{i}^{2}(k_z)- \mu_F \right)\right) \right)
\notag \\
&- \frac{L^3}{N_F\, \pi^2} \frac{m_e\,k_B\,T}{\hbar^2} \frac{1}{\beta} \sum_{i=1}^{j} \int\limits_{-0}^{\pi}\; dk_z f_2\left[-\exp\{-\beta[\alpha_{i}^{2}(k_z) - \mu_F] \} \right] \label{Eq-FF-KP}
\end{align}
%\end{small}
%%%
Expression \eqref{Eq-FF-KP} is plotted in Fig.~\ref{Fig-HF-KP}. Once again note that the IFG (red) curve coincides with the $P_{0_F} = 0$ (black-long-dashed) curve.
\break

\vspace*{-0.25cm}
\section{Condensation energy}

\begin{figure}[!ht]
\begin{center}
\includegraphics[width=0.45\textwidth]{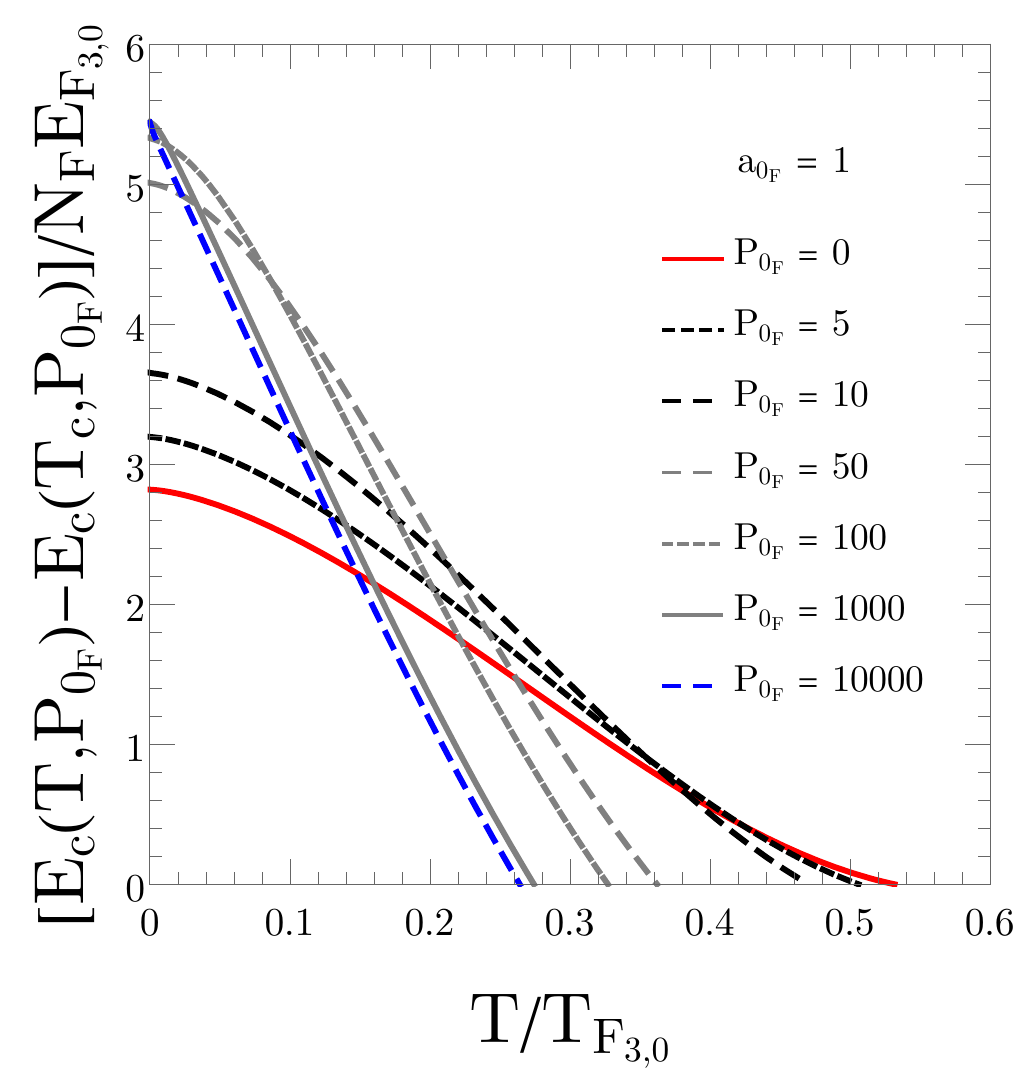}
\end{center}
\caption{Condensation energy $E_c(T)/N_F\,k_B\,T_F = [F_{N}(T) - 2F_{B}(T)]/N_F\,k_B\,T_F$, with $F_{F}(T)$  the fermion free energy and $ F_{B}(T)$ the boson free energy for different values of $P_{0_F}$ with $a_{0_F} = 1$. The condensation energy for $T=0$ increases as the strength impenetrability increases.}\label{Fig-EcFB}
\end{figure}

In order to obtain the condensation energy of the system, we assume that all of the $N_F$ fermions pair into $N_B = N_F/2$ CPs (bosons), so the condensation energy is given by 
\begin{eqnarray}
E_c(T) &=& F_{n}(T) - F_{s}(T)
\end{eqnarray}
where $F_{n} = F_F(T)$ is the Helmholtz free energy of the fermions and $F_s(T) = F_B(T) = F_{N_F/2}(T) $ is the Helmholtz free energy of the bosons. Thus, taking \eqref{Eq-FF-KP} and \eqref{Eq-FB-EF-KP} gives
\begin{widetext}
\begin{eqnarray}
%\begin{flalign}
\frac{E_c(T)}{N_F\,k_B\,T_F} &=& \frac{F_F(T)}{N_F\,k_B\,T_F} -\frac{2\,F_B(T)}{N_F\,k_B\,T_F}
\notag \\
&=& \frac{L^3}{N_F\, \pi ^2} \frac{m_e}{\hbar ^2} \frac{1}{\beta} \sum_{i=1}^{j} \int\limits_{0}^{\pi} \;dk_z~ \mu_F  \ln \left(1+ \exp \left(-\beta  \left(\alpha_{i}^{2}(k_z)- \mu_F \right)\right) \right)
+ \frac{L^3}{N_F\, \pi^2} \frac{m_e\,k_B\,T}{\hbar^2} \frac{1}{\beta} \sum_{i=1}^{j} \int\limits_{0}^{\pi}\; dk_z f_2\left[-\exp\{-\beta[\alpha_{i}^{2}(k_z) - \mu_F] \} \right]
\notag \\
&-& \frac{(e_0+\varepsilon _0)}{k_B\,T_F} + \frac{L^3}{2\,N_F\, \pi ^2 c^2}\frac{1}{\beta^2} \sum_{i=1}^{j} \int\limits_{0}^{\pi}\; dK_z ~ \left(\frac{e_0-\varepsilon _0-\mu_B}{k_B\,T_F} \right)
g_2\left(\exp\{-\beta [e_0 +\alpha_{i}^{2}(K_z) -\mu_B] \}\right)
\notag \\
&+& \frac{L^3}{2\,N_F\, \pi ^2 c^2 k_B\,T_F} \frac{1}{\beta^3} \sum_{i=1}^{j} \int\limits_{0}^{\pi}\; dK_z ~ g_3\left(\exp\{-\beta [e_0 +\alpha_{i}^{2}(K_z) -\mu_B] \} \right) \label{Ec-FB}
%\end{flalign}
\end{eqnarray}
\end{widetext}
This expression leads to the condensation energy curves which are plotted in Fig.\,\ref{Fig-EcFB} for several values of $P_{0_F}$ with $a_{0_F} =1$. As the plane impenetrability increases the condensation energy at $T=0$ increases, while at the same time the critical temperature decreases. Furthermore, for high values of $P_{0_F}$ the inflection of the curves at $T_c$, which is a signal of a second order transition \cite{GL1950}, vanishes, suggesting another kind of phase transition.
%%%%
%%%%

\section{Conclusions}
Using a Boson-Fermion formalism, we have calculated the condensation energy of a superconducting BEC inside a  multilayer structure, where the bosons are Cooper pairs coming from a fermion gas whose electrons interact attractively, assuming that all of the fermions are paired before turning off the interactions.
The multilayer structure is generated by applying an external Dirac-comb potential in the direction perpendicular to the layers.
Bosons have a linear energy-momentum relation in the directions parallel to the $x$-$y$ plane plus the energy coming from the Kronig-Penney energy relation perpendicular to that plane, while  fermions have a quadratic energy-momentum relation in the plane plus the corresponding energy due to the KP relation in the perpendicular direction.

We calculate several thermodynamic properties for bosons and fermions in multilayers  for various values of the layer impenetrability $P_{0_F}$ keeping the separation $a_{0_{F}}$ between planes constant. The critical temperature $T_c$ of the boson system decreases as $P_{0_F}$ increases, as already observed in \cite{SalasJLTP2010}. However, $a_{0_F}$ the distance between layers plays an important role, so for example, if the separation between planes increases, i.e. $a_{0_F} \to \infty$, we recover the IBG properties. 

The entropy of the boson system shows a phase transition, exhibits a change in its curvature precisely at critical temperature for each $P_{0_F}$. On the other hand, we observe that its ground state energy (boson chemical potential) increases, from the zero value corresponding to the free case, modulated by the influence of the multilayers, and that it crosses the horizontal axis at higher values as the plane impenetrability increases.

The internal energy over the temperature for the boson gas tends to $2.5$ (3D value) when $P_{0_F} =0 $,  while it tends to $2$ (2D value) when $P_{0_F} \to \infty$, both cases for the classical limit when $T/T_{F_{3,0}} \to \infty$, as expected. On the other hand, the internal energy over the temperature for the fermion gas tends to $1.5$ (3D value) with $P_{0_F} = 0$ and to $1$ (2D value) when $P_{0_F}\to  \infty$, as $T/T_{F_{3,0}} \to \infty$ also for both cases.
This behavior resembles a dimensional crossover shown in both, the entropy and internal energy, as the strength impenetrability increases. The obtained  curves  are qualitatively observed in layered high $T_c$ superconductors, such as cuprates and Fe-based.

Taking the condensation energy as the difference between the Helmholtz free energies of a normal state, formed by the fermion system in multilayers, minus the condensed state, formed by all the fermions paired into bosons, we observe that when the impenetrability strength of the planes increases, the condensation energy increases for all temperatures from zero to $T_c$, and qualitatively shows the same behavior of the experimental data. This feature will be very useful for calculating the condensation energy of high-$T_c$ superconductors, where we will be able to compare with experimental data by substituting the distance between planes $a_{0_F}$ by the cell size and associating the strength of the impenetrability $P_{0_{F}}$ with the size of the ion inside the structure of the lattice, which will be reported elsewhere.

\textbf{Acknowledgments}.
We thank partial support from DGAPA-PAPIIT-UNAM project grant IN114523. IC thanks to CONAHCyT-SECIHTI for the grant EPA1 \# 869450.\\

\appendix

\section*{Appendix A. Grand Thermodynamic potential of Bose and Fermi gases}

In general, the grand thermodynamic potential of ideal quantum gases is given by
\begin{equation}
	\Omega(T,L^d,\mu) = -\frac{\eta(2s+1)}{\beta} \sum_{\kappa} \ln \left[1+ \eta\,z\,\exp(-\beta\,\varepsilon_{\kappa}) \right]
\end{equation}
where $\beta = 1/k_B\,T$, $z \equiv \exp(\beta\mu)$ is the fugacity, $\mu$ is the chemical potential, $\eta=-1$ for bosons, $\eta=1$ for fermions and $s$ the spin of the particles. By taking the continuum case where the sums become integrals we get
\begin{equation}
	\Omega(T,L^d,\mu) = -\frac{\eta(2s+1)}{\beta^{d/2+1}} \left(\frac{m\,L^{2}}{2\pi\hbar^{2}} \right)^{d/2} Li_{d/2+1}[-\eta\,z]
\end{equation}
where
\begin{equation}
	-\eta\;Li_{\nu}[-\eta\,z] = \frac{1}{\Gamma(\nu)} \int_{0}^{\infty} dx \frac{x^{\nu -1}}{z^{-1}\exp(x)+\eta} \label{Eq-Polylog}
\end{equation}
which becomes either the Bose functions $g_{\nu}(z)$  or the Fermi functions $f_{\nu}(z)$. From here after we denote the fermion chemical potential as $\mu_F$ and the boson chemical potential as $\mu_B$. 

From the last two equations one gets the well known result of the thermodynamic potential for the the ideal Bose gas (IBG) with $d=3$ and $\eta=-1$
\begin{equation}
	\Omega_{IBG}(T,L^d,\mu_B) = \frac{1}{\beta^{5/2}} \left(\frac{m_b\,L^{2}}{2\pi\hbar^{2}} \right)^{3/2} g_{5/2}(z), \label{Omega-Bose3D}
\end{equation}
%with the boson mass $m_b = 2 m_e$ and $m_e$ the electron mass and 
while for the ideal Fermi gas (IFG) with $d=3$ and $\eta=1$
\begin{equation}
	\Omega_{IFG}(T,L^d,\mu_F) = -\frac{2}{\beta^{5/2}} \left(\frac{m_e\,L^{2}}{2\pi\hbar^{2}} \right)^{3/2} f_{5/2}(z). \label{Omega-Fermi3D}
\end{equation}

\section*{Appendix B. Boson Grand Thermodynamic Potential in multilayers}

For bosons the energy-momentum relation can be separated as $\varepsilon(K) = \varepsilon(K_x,K_y,K_z) = \varepsilon(K_{\rho}) + \varepsilon(K_z)$ with $\varepsilon(K_{\rho}) = 2E_{F_2} - \Delta_0 + c\, K_{\rho}$, $E_{F_2}$ the Fermi energy in 2D, $\Delta_0$ the energy (superconducting) gap of the Cooper pairs taken as bosons in 2D and $\varepsilon(K_z) = \alpha^{2}(K_z)/2$ is the KP energy relation for bosons cited above, thus
\begin{equation}
\Omega_B(T,L^3,\mu) = \frac{1}{\beta} \sum_{K} \ln \left[1-z\,\exp(-\beta\,[\varepsilon(K_{\rho}) + \varepsilon(K_z)]) \right].
\end{equation}
with $\mu_B$ the boson chemical potential. Using the logarithmic expansion
\begin{equation*}
\ln(1+x) = - \sum_{l=1}^{\infty} \frac{(-x)^l}{l}, \quad \text{for}~ |x|<1
\end{equation*}
and taking 
\begin{equation}
\sum_K \to  \left(\frac{L}{2\pi}\right)^{3} \int d^3\,K = \left(\frac{L}{2\pi}\right)^{3} \int_{K_z} dK_{z} \int_{0}^{2\pi} d\theta \int_{K_{\rho}}\,d\,K_{\rho} K_{\rho}
\end{equation}

Using cylindrical coordinates, we get 
\begin{small}
\begin{align*}
\Omega_B &= \Omega_{0} -\frac{1}{\beta}\left( \frac{L}{2\pi}\right)^{3} \sum_{l=1}^{\infty} \frac{\exp[\beta l \mu_B]}{l} \int_{K_z} dK_{z} \int_{0}^{2\pi} d\theta \int_{K_{\rho}} d\,K_{\rho} K_{\rho}
\notag \\
& \times  \exp[-\beta l \{ 2E_{F_2} - \Delta_0 + c\, K_{\rho} +\varepsilon(K_z) \}]
\notag \\
%%%
&= -\frac{1}{\beta} \left( \frac{L}{2\pi}\right)^{3} 2\pi \sum_{l=1}^{\infty} \frac{\exp[\beta l \mu_B]}{l}
\notag \\
&\times \int_{K_z} dK_{z} \exp[-\beta l \{2E_{F_2} - \Delta_0 + \varepsilon(K_z) \}]
\notag \\
&\times  \int_{K_{\rho}} d\,K_{\rho} K_{\rho} \exp[-\beta l \{ c\, K_{\rho} \}]
\end{align*}
\end{small}
where $\Omega_0$ is the thermodynamic potential of the ground state 
\begin{equation*}
\Omega_0 = \frac{1}{\beta} \ln \left[1- \exp\{-\beta (2E_{F_2} - \Delta_0 + \varepsilon_0 - \mu_B) \} \right]
\end{equation*}
and $\varepsilon_0$ is the ground state energy of the boson particles. By introducing a variable change with $u = \beta c K_{\rho}$, $du = \beta\,c\, dK_{\rho}$, and integrating, gives
\begin{eqnarray*}
%\begin{align*}
\Omega_B &=& \Omega_0
-\frac{L^3}{(2\,c\,\pi)^2} \frac{1}{(\beta)^3}  \int_{K_z} dK_{z} \sum_{l=1}^{\infty}
\frac{\exp[\beta l \mu_B]}{l^3}
\notag \\
&\times &  \exp[-\beta l \{2E_{F_2} - \Delta_0 +\varepsilon(K_z) \}]. 
%\end{align*}
\end{eqnarray*}
Let $e_0 = 2E_{F_2} - \Delta_0$, and rewriting
\begin{small}
\begin{eqnarray}
\Omega_B &=& \Omega_0 -\frac{L^3}{(2\pi)^2} \frac{1}{c^2\,\beta^3}  \int_{K_z} dK_{z} \sum_{l=1}^{\infty} \frac{\exp[-\beta l \{e_0 +\varepsilon(K_z) -\mu_B \}]}{l^3}
\notag \\
&=& \Omega_0 -\frac{L^3}{(2\pi)^2} \frac{1}{c^2\,\beta^3}  \int_{K_z} dK_{z}\; g_{3}\big( \exp[-\beta \{e_0 +\varepsilon(K_z) -\mu_B \}] \big) \notag
\end{eqnarray}
\end{small}
where $g_3(\exp[-\beta \{e_0 +\varepsilon(K_z) -\mu_B\}])$ is the Bose Function \cite{pathria} of third order. Finally using the KP relations for the boson gas we get
\begin{small}
%\begin{eqnarray}
\begin{align}
\Omega_B &= \Omega_0 -\frac{L^3}{2\,\pi^2} \frac{1}{c^2\,\beta^3}  \sum_{i=1}^{j} \int\limits_{0}^{\pi} dK_{z}\; g_{3}\big[ \exp(-\beta \{e_0 + \alpha_{i}^{2}(K_z)/2 -\mu_B \}) \big].
%\label{WBose}
\end{align}
%\end{eqnarray}
\end{small}
From the last equation we can calculate the thermodynamic properties, such as the chemical potential, critical temperature, entropy, internal energy, specific heat at constant volume and the Helmholtz free energy.

\section*{Appendix C. Fermion Grand Thermodynamic Potential in multilayers}

For the fermion system the energy-momentum relation can be separated as $\varepsilon(k) = \varepsilon(k_x,k_y,k_z) = \varepsilon(k_x+k_y) + \varepsilon(k_z)$. Substituting it in the thermodynamic potential for the Fermi gas 
\begin{equation}
\Omega_F(T,L^3,\mu_F) = -\frac{1}{\beta} \sum_{k} \ln \left[1-z\,\exp(-\beta\,[\varepsilon(k_x+k_y) + \varepsilon(k_z)]) \right]
\end{equation}
and after some algebra 
\begin{eqnarray}
\Omega_F &=& -\left(\frac{L}{2\pi}\right)^{3} \frac{1}{\beta}\;\sum_{l=1}^{\infty} \frac{\exp[-\beta\,l\,\mu_F]}{l} \int_{-\infty}^{\infty} dk_x~ \exp[-\beta\,l\tfrac{\hbar^2}{2m_e}k_x^2]
\notag \\
&\times & \int_{-\infty}^{\infty} dk_y~ \exp[-\beta\,l\tfrac{\hbar^2}{2m_e}k_y^2] \int_{-\infty}^{\infty} dk_z~ \exp[-\beta\,l \varepsilon(k_z)] \qquad
\end{eqnarray}

When we take the quadratic energy-momentum relation in the $x$-$y$ plane an evaluate the integrals we get
%\begin{eqnarray*}
\begin{align}
\Omega_F &= -\left(\frac{L}{2\pi}\right)^{3} \frac{1}{\beta}\;\sum_{l=1}^{\infty} \frac{\exp[-\beta\,l\,\mu_F]}{l}
\notag \\
&\times \left(\frac{2m_e\,\pi}{\beta\,l\,\hbar^{2}}\right) \int\limits_{-\infty}^{\infty} dk_z\; \exp[-\beta\,l\, \varepsilon(k_z)]
\notag \\
&= -\left(\frac{L}{2\pi}\right)^{3} \left(\frac{2m_e\,\pi}{\beta^2\,\hbar^{2}}\right) \; \int\limits_{-\infty}^{\infty} dk_z\; \sum_{l=1}^{\infty} \frac{\exp[\beta\,l\,(\mu-\varepsilon(k_z))]}{l^{2}} \quad \Rightarrow
\notag \\
\Omega_F &= -\frac{L^3}{2\,\pi^2} \left(\frac{m_e}{\beta^2\,\hbar^{2}}\right) \;\int\limits_{-\infty}^{\infty} dk_z\; f_{2}\big[\exp[-\beta\,(\varepsilon(k_z)-\mu_F)]\big]
\end{align}
%\end{eqnarray*}
Taking the KP relations cited above finally gives
%%\begin{eqnarray}
\begin{align}
\Omega_F &= -\frac{L^3}{\pi^2} \left(\frac{m_e}{\beta^2\,\hbar^{2}}\right) \sum_{i=1}^{j}\int_{0}^{\pi} dk_z\; f_{2}\big[\exp[-\beta\,(\alpha_{i}^{2}(k_z)-\mu_F)]\big].
%\label{WFermi}
\end{align}
%%\end{eqnarray}

\end{document}